\documentclass[utf8]{mystylepaper} 

\setcitestyle{square}
\usepackage{url,hyperref,lineno,microtype,subcaption}
\usepackage[onehalfspacing]{setspace}
\usepackage{siunitx}

\def\keyFont{\fontsize{8}{11}\helveticabold }
\def\firstAuthorLast{Marzioni {et~al.}}
\def\Authors{Francesco Marzioni\,$^{1,2,4,\dagger}$, Francesco Rasponi\,$^{1,\dagger}$, Paolo Piergentili\,$^{1,2}$, Riccardo Natali\,$^{1,2}$, Giovanni Di Giuseppe\,$^{1,2}$ and David Vitali\,$^{1,2,3,*}$}
\def\Address{$^{1}$School of Science and Technology, Physics Division, University of Camerino, I-62032 Camerino (MC), Italy \\
$^{2}$INFN, Sezione di Perugia, Italy \\
$^{3}$CNR-INO, L.go Enrico Fermi 6, I-50125 Firenze, Italy \\
$^{4}$Department of Physics, University of Naples “Federico II”, I-80126 Napoli, Italy\\
$\dagger$ These authors contributed equally to this work}
\def\corrAuthor{David Vitali}
\def\corrEmail{david.vitali@unicam.it}

\begin{document}
\onecolumn
\firstpage{1}

\title {Amplitude and phase noise in Two-membrane cavity optomechanics}

\author[\firstAuthorLast ]{\Authors} 
\address{} 
\correspondance{} 
\extraAuth{}

\maketitle

\begin{abstract}

Cavity optomechanics is a suitable field to explore quantum effects on macroscopic objects, and to develop quantum technologies applications.
A perfect control on the laser noises is required to operate the system in such extreme conditions, necessary to reach the quantum regime.
In this paper we consider a Fabry-Per\'ot cavity, driven by two laser fields, with two partially reflective SiN membranes inside it.
We describe the effects of amplitude and phase noise on the laser introducing two additional noise terms in the Langevin equations of the system's dynamics.
Experimentally, we add an artificial source of noise on the laser.
We calibrate the intensity of the noise we inject into the system, and we check the validity of the theoretical model.
This procedure provides an accurate description of the effects of a noisy laser in the optomechanical setup, and it allows to quantify the amount of noise.
\section{}
\tiny
 \keyFont{ \section{Keywords:} cavity optomechanics, radiation pressure, membranes, laser noise, calibration} 
\end{abstract}

\section{Introduction}

In cavity optomechanics a number of mechanical degrees of freedom is coupled by a dispersive parametric coupling, typically due to the radiation pressure interaction, to one or more driven modes of an optical cavity~\cite{Aspelmeyer2014}.
Such a configuration can be realized in a large variety of platforms, with different cavity geometries, as well as very different kind of mechanical modes, ranging from vibrational modes of 1D, 2D or 3D systems, to the center-of-mass motion of trapped nanoparticles, to bulk and surface acoustic waves of properly designed materials.
Despite their macroscopic nature, these devices can now be operated in a quantum regime, for a number of relevant applications in quantum technologies, and for answering fundamental physics questions.
For example, entangled states of radiation modes~\cite{Barzanjeh2019,Chen2020}, of mechanical modes~\cite{Ockeloen2018,Kotler2021} and hybrid optomechanical entanglement~\cite{Palomaki:2013aa} have been generated, as well as squeezed states of optical~\cite{Brooks:2012aa,Safavi-Naeini:2013aa,Purdy:2013aa} and mechanical~\cite{Pirkkalainen:2015aa} modes.
Detection of displacements and forces below the standard quantum limit (SQL) has been demonstrated \cite{Mason:2019aa}, and also microwave-optical transduction with added noise very close to the quantum limit has been demonstrated \cite{Brubaker:2022aa,Seis:2022aa}.

In all these experiments the required effective interaction between the mechanical and the cavity modes is obtained by engineering the amplitudes and the frequencies of the driving sources, which have to be phase-locked and stabilized as much as possible.
In fact, as it can be easily expected, both amplitude and phase noise of the laser driving is detrimental for any quantum effects, as it has been already theoretically suggested~\cite{Rabl2009,Abdi2011} and experimentally verified.
In fact, analyses of the effect of laser phase noise on cooling of the mechanical mode appeared in Refs.~\cite{Schliesser2008,Jayich:2012ab,Meyer2019}, while its effect on squeezing was analysed in Ref.~\cite{Pontin2014}.
On the one hand one has to minimize as much as possible these technical noise sources, for example by using filter cavities or moving to higher mechanical resonance frequencies where these noises become less relevant; on the other hand it would be useful to develop a procedure for modelling and quantifying in a simple way the effect of these noise.
Here we provide such a procedure, by considering a multimode cavity optomechanical system in which two SiN membranes are placed within a Fabry-Per\'ot cavity and driven by two laser fields, which has been considered for a large number of physical processes, such as synchronization of mechanical modes~\cite{Sheng2020,Piergentili2020}, mechanical state swapping~\cite{Weaver2017}, heat transfer~\cite{Yang2020}, cooperativity competition~\cite{deJong2022}, and enhancement of single-photon optomechanical coupling~\cite{Piergentili2018,Li:2016aa}.
Here we artificially add amplitude and phase noise to one of the driving field, the pump beam controlling the interaction in the linearized regime of cavity optomechanics~\cite{Aspelmeyer2014}, and show how one can calibrate and quantify the corresponding noise spectra, and describe it in terms of the size of the noise spectra of additional amplitude and phase noise terms in the Langevin equations describing the dynamics.

The calibration and modelization of the laser noise presented here is general and it can be applied to a generic optomechanical systems. For its experimental verification we consider here the so-called ``membrane-in-the-middle''~\cite{Thompson:2008aa,Wilson:2009aa} configuration of cavity optomechanics, which has been employed by many groups since 2008~\cite{Jayich:2012ab,Sheng2020,Piergentili2020,Weaver2017,Yang2020,Thompson:2008aa,Wilson:2009aa}, and presents many advantages. In fact, thin semitransparent Si$_3$N$_4$ membranes are commercially available, and presents a very high mechanical quality factor due its intrinsic high-stress~\cite{Southworth:2009aa}. Moreover, they are characterized by a negligible absorption at optical wavelengths~\cite{Serra:2016ab}, so that, if they are placed near the waist of an optical Fabry-Per\'ot cavity, losses due to scattering are negligible and the cavity decay rate is mostly determined by the empty cavity finesse only~\cite{Jayich:2012ab,Thompson:2008aa,Wilson:2009aa,Serra:2016ab}. Furthermore, the single-photon optomechanical coupling with a given optical cavity mode can be fine-tuned by controlling the longitudinal and transversal position of the membrane within the cavity~\cite{Thompson:2008aa,Wilson:2009aa,Biancofiore:2011aa}. These facts make membrane-in-the-middle optomechanical setups particularly suitable for operating in the resolved sideband regime, and for reaching a very large optomechanical cooperativity, which are fundamental conditions for the realization and manipulation of quantum states of the cavity mode and of the mechanical resonator~\cite{Aspelmeyer2014}, and witnessed, for example by the results of Refs.~\cite{Chen2020,Purdy:2013aa,Mason:2019aa,Brubaker:2022aa}. Recent advances have shown that silicon nitride membranes are particularly suitable for further engineering and suppression of clamping losses, either via
the use of on-chip seismic filtering stage~\cite{Borrielli:2016aa}, or of phononic bandgap crystal designs~\cite{Tsaturyan:2017aa}, which have allowed very large mechanical quality factors, even at room temperature~\cite{Tsaturyan:2017aa,Serra:2021aa}.

The paper is organized as follows.
In Sec.~II we present the model Hamiltonian and we show how one can modify the standard QLE treatment in order to include the effects of laser noise.
In Sec.~III we describe the experimental setup, and we show the detected noise spectra, either the output spectrum in transmission and the homodyne detection of the mechanical motion of the two membranes. In addition we show how one can reconstruct and model the effect of the amplitude and phase noise of the driving fields.
Sec.~IV is for concluding remarks.

\section{The model}

\begin{figure}
\centering
\includegraphics[scale=0.71]{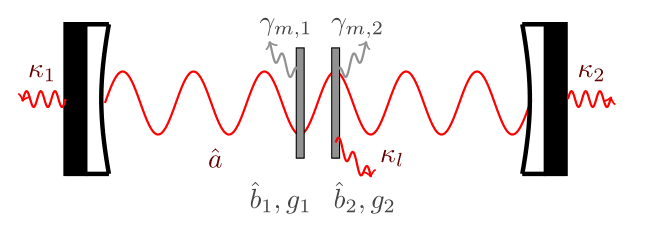}
\caption{Two SiN membranes within a Fabry-Per\'ot cavity. The two mechanical resonators $\hat{b}_1$, $\hat{b}_2$, having masses $m_1$, $m_2$, resonance frequencies $\omega_{m,1}$, $\omega_{m,2}$ and damping rates $\gamma_{m,1}$, $\gamma_{m,2}$ interact with an optical mode $\hat{a}$. The cavity is characterized by decay rate $\kappa_1$ for the input port, $\kappa_2$ for the output and $\kappa_l$ that takes into account losses due to absorption or scattering.}
\label{fig:model}
\end{figure}

The system is composed by two mechanical resonators which are effectively coupled through two driven optical cavity modes, as sketched in Fig.~\ref{fig:model}, the pump mode that is responsible for the optomechanical interaction, and the probe mode to perform the spectral analysis.
The Hamiltonian hence includes several components:
\begin{align}
 \hat{H}/\hbar=\sum_{k=1,2}\Bigl\{&\omega_{c,k}\hat{a_k}^\dagger\hat{a_k}+\sum_{j=1,2}\left[\omega_{m,j}\hat{b}_j^\dagger\hat{b}_j-g_{jk}^{(0)}\hat{a_k}^\dagger\hat{a_k}(\hat{b}_j+\hat{b}_j^\dagger)\right] \\
 \nonumber &+ iE_k(t)(\hat{a_k}^\dagger e^{-i[\omega_{L,k}t+\phi_k(t)]}-\hat{a}e^{i[\omega_{L,k}t+\phi_k(t)]})\Bigr\}\label{ham1},
\end{align}
the first term describes the energy of the cavity modes represented by bosonic annihilation operator $\hat{a_k}$ $([\hat{a_k},\hat{a_k}^\dagger]=1)$ at frequency $\omega_{c,k}$, where we consider $\hat{a_1}$ the operator related to the probe mode and $\hat{a_2}$ to the pump mode; the second term takes into account the energies of the two mechanical resonators with annihilation operators $\hat{b}_j$ $([\hat{b}_j,\hat{b}_j^\dagger]=1)$ at frequency $\omega_{m,j}$ for the $j$-th membrane and the optomechanical interactions, that arise from the dependence of the optical mode frequencies on the positions of the mechanical resonators, and characterized by single photon coupling rate $g_{jk}^{(0)}=-\sqrt{\hbar/2m_j\omega_{m,j}}\cdot\partial_{x_j}\omega_{c,k}(x_j)$. The third and last describes the laser driving fields at frequencies $\omega_{L,k}$.
We consider general driving fields which are affected by fluctuations, both in amplitude and phase.
The amplitude is related to the number of photons that enter into the cavity every second and can be written as $E_k(t)=\bar{E_k}+\epsilon_k(t)$.
The mean value corresponds to $\bar{E_k}=\sqrt{P_k\kappa_{1k}/\hbar\omega_{L,k}}$, being $P_k$ the laser power of the $k$-th optical mode, $\kappa_{1k}$ the cavity decay rate for the input mirror, while the time dependent term $\epsilon_k(t)$ represents the real, zero-mean amplitude fluctuations of the $k$-th driving field.
We considered $\bar{E_k}$ real, which means that we are choosing the laser driving field as phase reference for the $k$-th optical field.
Similarly, we call $\omega_{L,k}$ the average laser frequency, and $\phi_k(t)$ denotes the zero mean fluctuating phase.
In order to remove the explicit time dependence from the Hamiltonian and consequently from the Langevin equations, one can choose the frame rotating at the laser frequency.
Our case involves the exploitation of a fluctuating frequency laser so we pose in the randomly rotating frame at instantaneous frequency $\omega_{L,k}+\dot{\phi}_k(t)$.
One obtains a new form for the Hamiltonian devoid of the time dependent exponentials of the laser driving term:
\begin{equation}\label{ham_rot}
 \hat{H}/\hbar=\sum_{k=1,2}\Bigl\{(\Delta^{(0)}_k-\dot{\phi}_k)\hat{a_k}^\dagger\hat{a_k}+\sum_{j=1,2}\left[\omega_{m,j}\hat{b}_j^\dagger\hat{b}_j-g_{jk}^{(0)}\hat{a_k}^\dagger\hat{a_k}(\hat{b}_j+\hat{b}_j^\dagger)\right]+iE_k(t)(\hat{a_k}^\dagger-\hat{a_k})\Bigr\},
\end{equation}
where we defined the detuning $\Delta_k^{(0)}=\omega_{c,k}-\omega_{L,k}$.
From the latter expression for the Hamiltonian, we are able to derive the quantum Langevin equations for the optical and mechanical modes as:
\begin{align}
\dot{\hat{a}}_k=&-\left[\frac{\kappa_k}{2}+i(\Delta_k^{(0)}-\dot{\phi}_k)\right]\hat{a_k}+i\sum_{j=1,2}g_{jk}^{(0)}(\hat{b}_j+\hat{b}_j^\dagger)\hat{a_k}+E_k+\sqrt{\kappa_{1k}}(\hat{a}_{in,1k}+\epsilon_k)+\sqrt{\kappa_{2k}}\hat{a}_{in,2k},\label{lang_a}\\
\dot{\hat{b}}_j=&-\left(\frac{\gamma_{m,j}}{2}+i\omega_{m,j}\right)\hat{b}_j+i\sum_{k=1,2}g_{jk}^{(0)}\hat{a_k}^\dagger\hat{a_k}+\sqrt{\gamma_{m,j}}\hat{b}_{in,j}.\label{lang_b}
\end{align}
The Langevin equation for the optical mode takes into account the optical noise operators related to the input mirror $\hat{a}_{in,1k}$, characterized by decay rate $\kappa_{1k}$, and to the output mirror $\hat{a}_{in,2k}$ which looses photons at rate $\kappa_{2k}$, meaning that the whole cavity has a total decay rate $\kappa_k=\kappa_{1k}+\kappa_{2k}+\kappa_{l}$, where $\kappa_{l}$ takes into account the other losses mechanisms.
Note that the laser amplitude noise is added to the optical noise: the system is indeed affected by the vacuum optical noise as well as the laser amplitude fluctuations.
Differently, the phase frequency noise affects the optical mode acting as a multiplicative noise, similarly to the position fluctuations of the mechanical resonators that arise because of the optomechanical interaction.
Technical noises effects can be negligible or relevant depending on the quality of the laser itself.
In fact, as will be shortly outlined, the features of the laser in connection with the frequency region of interest determine the significance of the laser noises.
The Langevin equation for the $j$-th mechanical mode is governed by thermal noise $\hat{b}_{in,j}$, and damping rate $\gamma_{m,j}$, and shows an optomechanical interaction proportional to the number of intracavity photons.
We consider Markovian correlation functions for the optical and mechanical input noise operators:
\begin{align}
 \langle\hat{a}_{in,lk}(t)\hat{a}_{in,lk'}^\dagger(t')\rangle&=\delta_{kk'}\delta(t-t'),\label{a_adag}\\
 \langle\hat{b}_{in,j}^\dagger(t)\hat{b}_{in,j'}(t')\rangle&=n_{j}^{th}\delta_{jj'}\delta(t-t'),\label{bdag_b}\\
 \langle\hat{b}_{in,j}(t)\hat{b}_{in,j'}^\dagger(t')\rangle&=(n_{j}^{th}+1)\delta_{jj'}\delta(t-t'),\label{b_bdag}
\end{align}
where the $l$-index refers to the input or output port of interest.
Both optical and mechanical noises are thermal, hence proportional to the thermal bosons number given by the Bose-Einstein statistics $n^{th}=(e^{\hbar \omega/K_BT}-1)^{-1}$.
Specifically, at room temperature, and at the cavity frequency $\hbar\omega_{c,k}/K_BT\gg1$, so that the optical noise can be treated as vacuum noise, the only relevant correlation is Eq.~\eqref{a_adag}; differently, at the mechanical resonant frequencies one can consider $n_j^{th}\simeq K_BT_j/\hbar\omega_j$.
The correlation functions for the amplitude and phase noises have the following expressions:
\begin{align}
 \langle\epsilon_k(t)\epsilon_k(t')\rangle&=\Gamma_{\epsilon,k}\gamma_{\epsilon,k} e^{-\gamma_{\epsilon,k}|t-t'|},\label{ee}\\
 \langle\dot{\phi}_k(t)\dot{\phi}_k(t')\rangle&=\Gamma_{L,k}\gamma_{\phi,k} e^{-\gamma_{\phi,k}|t-t'|},\label{pp}
\end{align}
where $\Gamma_{\epsilon,k}$ is a dimensionless parameter that quantifies the intensity of the laser amplitude fluctuations and $\gamma_{\epsilon,k}$ corresponds to the bandwidth of the amplitude noise spectrum; analogously $\Gamma_{L,k}$ denotes the strength of the laser phase noise but has dimension of a frequency and $\gamma_{\phi,k}$ represents the bandwidth of the phase noise.
More precisely, $\Gamma_{L,k}$ stands for the linewidth of the laser that characterizes the laser spectrum itself, and typical values span in the range $\sim1-100$ kHz \cite{Abdi2011}.
Performing the Fourier transform of such correlations one obtains the following spectra for the noises:
\begin{align}\label{eq:Gamma_noise}
 S_{\epsilon,k}(\omega)=\Gamma_{\epsilon,k}\frac{2\gamma_{\epsilon,k}^2}{\gamma_{\epsilon,k}^2+\omega^2},\\
 S_{\phi,k}(\omega)=\Gamma_{L,k}\frac{2\gamma_{\phi,k}^2}{\gamma_{\phi,k}^2+\omega^2}.
\end{align}
Therefore we consider the amplitude and phase noises as colored noises with a Lorentzian spectrum, differently from the Markovian vacuum optical and thermal mechanical noises which have a flat frequency noise spectrum.
Indeed a flat spectrum tends to overestimate the effect of laser noises \cite{Rabl2009}.
However, we note that when $\gamma_{\epsilon}\gg\omega_{m,j}$, or $\gamma_{\phi}\gg\omega_{m,j}$, we recover a flat frequency spectrum situation.

\subsection{Linearized quantum Langevin equations}

We focus on the stationary state of the system, and our purpose is to investigate the dynamics of the membranes around the equilibrium positions due to the several noise sources.
Hence we consider the annihiliation operators of the quantum Langevin equations as composed by a mean amplitude term and a fluctuation term around that value, so that the optical annihilation operator can be written as $\hat{a}=\alpha+\delta{\hat{a}}$ and the mechanical one results $\hat{b}=\beta+\delta{\hat{b}}$.
Disregarding the noises, by inserting this decomposition for the operators into Eqs.~\eqref{lang_a}, \eqref{lang_b} and retaining the zero-th order terms, we get the steady state values:
\begin{align}
 \alpha_k=&\frac{E_k}{\kappa_k/2+i\Delta_k},\\
 \bar{x}_j=\beta_j+\beta_j^*=&\frac{2\sum_{k}g_{jk}^{(0)}|\alpha|^2\omega_{m,j}}{\gamma_{m,j}^2/4+\omega_{m,j}^2}\simeq\frac{2\sum_{k}g_{jk}^{(0)}|\alpha|^2}{\omega_{m,j}},
\end{align}
here we introduced a new detuning $\Delta_k=\Delta_k^{(0)}-\sum_jg_{jk}^{(0)}\bar{x}_j$ which is influenced by the stationary position of both mechanical resonators, and an effective coupling rate $g_{jk}=g_{jk}^{(0)}\alpha_k$.
The final expression for the mean position of each mechanical resonator is valid in the case of damping rate much lower than the resonant frequency $\gamma_{m,j}\ll\omega_{m,j}$, a condition satisfied in optomechanical devices.
Now, taking into account the first order fluctuation terms, we get the linearized quantum Langevin equations for the annihilation operators:
\begin{align}
\delta\dot{\hat{a}}_k=&-\left(\frac{\kappa_k}{2}+i\Delta_k\right)\delta\hat{a}+i\sum_{j=1,2}g_{jk}(\delta\hat{b}_j+\delta\hat{b}_j^\dagger)+\sqrt{\kappa_{1k}}(\hat{a}_{in,{1k}}+\epsilon_k)+i\alpha_k\dot{\phi}_k+\sqrt{\kappa_{2k}}\hat{a}_{in,{2k}}\label{lang_a1}\\
\delta\dot{\hat{b}}_j=&-\left(\frac{\gamma_{m,j}}{2}+i\omega_{m,j}\right)\delta\hat{b}_j+i\sum_{k=1,2}g_{jk}^{(0)}(\alpha_k^*\delta \hat{a}_k+\alpha_k\delta\hat{a}^\dagger_k)+\sqrt{\gamma_{m,j}}\hat{b}_{in,j}\label{lang_b1} .
\end{align}
We have omitted all the second order terms, which are negligible when $|\alpha|\gg1$.
From these equations it is evident that in the linearized regime the effect of laser phase noise can be more relevant than that of amplitude noise, since it is multiplied by the intracavity amplitude $\alpha$.
Switching to the frequency domain by performing the Fourier transform of Eqs.\eqref{lang_a1}, \eqref{lang_b1}, we obtain:
\begin{align}
\left[\chi_{c,k}(\omega)\right]^{-1}\hat{a}_k&=i\sum_{j=1,2}g_{jk}(\hat{b}_j+\hat{b}_j^\dagger)+\sqrt{\kappa_{1k}}(\hat{a}_{in,{1k}}+\epsilon_k)+i\alpha_k\dot{\phi}_k+\sqrt{\kappa_{2k}}\hat{a}_{in,2k},\label{langf_a}\\
\left[\chi_{c,k}(-\omega)^{*}\right]^{-1}\hat{a}_k^\dagger&=-i\sum_{j=1,2}g_{jk}^*(\hat{b}_j+\hat{b}_j^\dagger)+\sqrt{\kappa_{1k}}(\hat{a}_{in,1k}^\dagger+\epsilon_k)-i\alpha_k^*\dot{\phi}_k+\sqrt{\kappa_{2k}}\hat{a}_{in,2k}^\dagger,\label{langf_a_dag}\\
\left[\chi_{m,j}(\omega)\right]^{-1}\hat{b}_j&=i\sum_{k=1,2}(g_{jk}^*\hat{a}_k+g_{jk}\hat{a}_k^\dagger)+\sqrt{\gamma_{m,j}}\hat{b}_{in,j},\label{langf_b}\\
\left[\chi_{m,j}(-\omega)^{*}\right]^{-1}\hat{b}^\dagger_j&=-i\sum_{k=1,2}(g_{jk}^*\hat{a}_k+g_{jk}\hat{a}_k^\dagger)+\sqrt{\gamma_{m,j}}\hat{b}^\dagger_{in,j}.\label{langf_b_dag}
\end{align}
The previous equations refer to the fluctuation operators, and we dropped out the fluctuation symbol $\delta$ for convenience.
We introduced the optical and mechanical bare susceptibilities which result:
\begin{align}
 \left[\chi_{c,k}(\omega)\right]^{-1}=&\frac{\kappa_k}{2}-i(\omega-\Delta_k),\label{susc_c}\\
 \left[\chi_{m,j}(\omega)\right]^{-1}=&\frac{\gamma_{m,j}}{2}-i(\omega-\omega_{m,j})\label{susc_j}.
\end{align}
In particular, for the probe field we consider from now on a detuning $\Delta_1=0$.
By inserting Eqs.~\eqref{langf_a}, \eqref{langf_a_dag} into Eq.~\eqref{langf_b} one obtains the equation for one mechanical annihilation operator as a function of the other mechanical operators and noises:
\begin{align}
\left[\chi_{m,j}'(\omega)\right]^{-1}\hat{b}_j=&-i\sum_{k=1,2}g_{jk}^{(0)}\sigma_k(\omega)\bigl[g_{jk}^{(0)}\hat{b}_j^\dagger+g_{3-jk}^{(0)}(\hat{b}_{3-j}+\hat{b}_{3-j}^\dagger)+\dot{\phi}_k\bigr]\\
\nonumber&+i\sum_{k=1,2}g_{jk}^{(0)}\Bigl\{\alpha^*_k\chi_{c,k}(\omega)\bigl[\sqrt{\kappa_{1k}}(\hat{a}_{in,1k}+\epsilon_k)+\sqrt{\kappa_{2k}}\hat{a}_{in,2k}\bigr] \Bigr.\\
\nonumber&\Bigl.+\alpha_k\chi_{c,k}^*(-\omega)\bigl[\sqrt{\kappa_{1k}}(\hat{a}_{in,1k}^\dagger+\epsilon_k)+\sqrt{\kappa_{2k}}\hat{a}_{in,2k}^\dagger\bigr]\Bigr\}+\sqrt{\gamma_{m,j}}\hat{b}_{in,j},
\end{align}
where we have defined $\sigma_k(\omega)=i|\alpha_k|^2\bigl[\chi_{c,k}^*(-\omega)-\chi_{c,k}(\omega)\bigr]$ and $\left[\chi_{m,j}'(\omega)\right]^{-1}=\left[\chi_{m,j}(\omega)\right]^{-1}+i\sum_k\sigma_{jk}(\omega)$, being $\sigma_{jk}(\omega)=g_{jk}^{(0)2}\sigma_k(\omega)$.
We now choose to work in rotating wave approximation, valid in red detuned regime ($\Delta>0$), weak coupling, and resolved sideband limit ($g_j < \kappa\leq \omega_{m,j}$), hence we neglect the counter-rotating mechanical terms $\hat{b}^\dagger$ and obtain the following equation for the $j$-th mechanical annihilation operator, which depends solely on the noises terms:
\begin{align}
 &\label{b_rwa}\left[\chi^{rwa}_{m,j}(\omega)\right]^{-1}\hat{b}_j=-i\sum_{k=1,2}g_{jk}^{(0)}\sigma_k(\omega)\left[-i\sum_{k'=1,2}g_{3-jk}^{(0)}g_{3-jk'}^{(0)}\sigma_{k'}(\omega)\chi'_{m,3-j}(\omega)\dot{\phi}_{k'}+\dot{\phi}_{k}\right]\\
 \nonumber &+\sum_{k=1,2}g_{jk}^{(0)}\Biggl\{\sum_{k'=1,2}\sigma_k(\omega)g_{3-jk}^{(0)}g_{3-jk'}^{(0)}\chi'_{m,3-j}(\omega)\Bigl[\alpha_{k'}^*\chi_{c,k'}(\omega)\bigl(\sqrt{\kappa_{1k'}}(\hat{a}_{in,1k'}+\epsilon_{k'}\bigl)+\sqrt{\kappa_{2k'}}\hat{a}_{in,2k'})\\
 \nonumber&+\alpha_{k'}\chi_{c,k'}^*(-\omega)\bigl(\sqrt{\kappa_{1k'}}(\hat{a}_{in,1k'}^\dagger+\epsilon_{k'})+\sqrt{\kappa_{2k'}}\hat{a}_{in,2k'}^\dagger\bigr)\Bigr]+i\Bigl[\alpha_{k}^*\chi_{c,k}(\omega)\bigl(\sqrt{\kappa_{1k}}(\hat{a}_{in,1k}+\epsilon_{k})+\sqrt{\kappa_{2k}}\hat{a}_{in,2k}\bigr)\\
 \nonumber&\hspace{2.7cm}+\alpha_{k}\chi_{c,k}^*(-\omega)\bigl(\sqrt{\kappa_{1k}}(\hat{a}_{in,1k}^\dagger+\epsilon_{k})+\sqrt{\kappa_{2k}}\hat{a}_{in,2k}^\dagger\bigr)\Bigr]\Biggr\}\\
 \nonumber&\hspace{2.7cm}-i\sum_{k=1,2}g_{jk}^{(0)}g_{3-jk}^{(0)}\sigma_k(\omega)\chi'_{m,3-j}(\omega)\sqrt{\gamma_{m,3-j}}\hat{b}_{in,3-j}+\sqrt{\gamma_{m,j}}\hat{b}_{in,j},
\end{align}
where the effective susceptibility in rotating wave approximation results:
\begin{equation}
 \left[\label{chi_rwa}\chi^{rwa}_{m,j}(\omega)\right]^{-1}=\left[\chi_{m,j}'(\omega)\right]^{-1}+\sum_{k,k'=1,2}g^{(0)}_{j,k}g^{(0)}_{j,k'}\sigma_k\sigma_{k'}g^{(0)}_{3-j,k}g^{(0)}_{3-j,k'}\chi_{3-j}(\omega).
\end{equation}
The first term of Eq.~\eqref{b_rwa} takes into account the phase noise contributions for both the probe and pump laser beams, the second one refers to the vacuum optical and amplitude laser noises for the input and output ports, while the latter term is related to the mechanical thermal noise.

\subsection{Output quadrature spectrum}

We detect the noise spectrum of the general amplitude quadrature $\hat{X}_k=i(\hat{a}^\dagger_ke^{i\varphi}+\hat{a}_ke^{-i\varphi})/\sqrt{2}$.
In particular, if we are interested in monitoring the pump field signal (which we denoted as optical mode 2) in transmission (hence measured from the output port 2), using the input-output relation we get $\hat{X}_{out,22}=\sqrt{\kappa_{22}}\hat{X}_2-\hat{X}_{in,22}$, where $\hat{X}_{in,22}=(\hat{a}^\dagger_{in,22}e^{i\varphi}+\hat{a}_{in,22}e^{-i\varphi})/\sqrt{2}$.
The expression from the output amplitude quadrature results:
\begin{align}
 \label{X_out}\hat{X}_{out,22}=\sqrt{\frac{\kappa_{22}}{2}}\Biggl[&-i\Bigl(\alpha_2^*\chi_{c,2}^*(-\omega)e^{i\varphi}-\alpha_2\chi_{c,2}(\omega)e^{-i\varphi}\Bigr)\biggl(\sum_{j=1,2}g_{j2}^{(0)}(\hat{b}_j+\hat{b}_j^\dagger)+\dot{\phi}_2\biggr)\\
 \nonumber &+\sqrt{\kappa_{12}}\Bigl[\chi_{c,2}^*(-\omega)e^{i\varphi}(\hat{a}^\dagger_{in,12}+\epsilon_2)+\chi_{c,2}(\omega)e^{-i\varphi}(\hat{a}_{in,12}+\epsilon_2)\Bigr]\\
 \nonumber &+\sqrt{\kappa_{22}}\Bigl[\Bigl(\chi_{c,2}^*(-\omega)-\frac{1}{\kappa_{22}}\Bigl)e^{i\varphi}\hat{a}^\dagger_{in,22}+\Bigl(\chi_{c,2}(\omega)-\frac{1}{\kappa_{22}}\Bigl)e^{-i\varphi}\hat{a}_{in,22}\Bigr]\Biggr],
\end{align}
so that, inserting Eq.~\eqref{b_rwa} for the annihilation and creation mechanical modes in Eq.~\eqref{X_out}, one gets the full, but cumbersome, expression in term of the noises.
The related spectrum can be calculated as $S_{X_{out}}(\omega)=\int_{-\infty}^\infty\langle\hat{X}_{out}(\omega)\hat{X}_{out}(\omega')\rangle d\omega'$ (see \cite{Aspelmeyer2014}). Hence, in order to derive the expression for the auto-correlation of the output optical quadrature, one needs to insert in the calculation the auto-correlation for each noise source, which are reported in Eqs.~\eqref{a_adag}-\eqref{pp} in time domain. However, since we are working in frequency domain, it is necessary to perform the Fourier transform of the such quantities.

Similarly the general phase quadrature $\hat{Y}_k=i(\hat{a}^\dagger_ke^{i\varphi}-\hat{a}_ke^{-i\varphi})/\sqrt{2}$ can be obtained for the probe field signal (which we denoted as optical mode 1) in reflection (hence measured from the output port 1).
From the input-output relation we have $\hat{Y}_{out,11}=\sqrt{\kappa_{11}}\hat{Y}_1-\hat{Y}_{in,11}$, where $\hat{Y}_{in,11}=i\left[(\hat{a}^\dagger_{in,11}+\epsilon_1)e^{i\varphi}-(\hat{a}_{in,11}+\epsilon_1)e^{-i\varphi}\right]/\sqrt{2}$. The expression results:
\begin{align}
 \label{Y_out}\hat{Y}_{out,11}=i\sqrt{\frac{\kappa_{11}}{2}}\Biggl[&-i\Bigl(\alpha_1^*\chi_{c,1}^*(-\omega)e^{i\varphi}+\alpha_1\chi_{c,1}(\omega)e^{-i\varphi}\Bigr)\biggl(\sum_{j=1,2}g_{j1}^{(0)}(\hat{b}_j+\hat{b}_j^\dagger)+\dot{\phi}_1\biggr)\\
 \nonumber &+\sqrt{\kappa_{11}}\Bigl[\Bigl(\chi_{c,1}^*(-\omega)-\frac{1}{\kappa_{11}}\Bigr)e^{i\varphi}(\hat{a}^\dagger_{in,11}+\epsilon_1)-\Bigl(\chi_{c,1}(\omega)-\frac{1}{\kappa_{11}}\Bigr)e^{-i\varphi}(\hat{a}_{in,11}+\epsilon_1)\Bigr]\\
 \nonumber &+\sqrt{\kappa_{21}}\Bigl(\chi_{c,1}^*(-\omega)e^{i\varphi}\hat{a}^\dagger_{in,21}-\chi_{c,1}(\omega)e^{-i\varphi}\hat{a}_{in,21}\Bigr)\Biggr].
\end{align}
As above one can derive the expression in terms of the noises and the related spectrum.

\section{Material and Methods}
\subsection{Description of the experimental setup}

\begin{figure}[htbp]
\begin{center}
\includegraphics[scale=0.83]{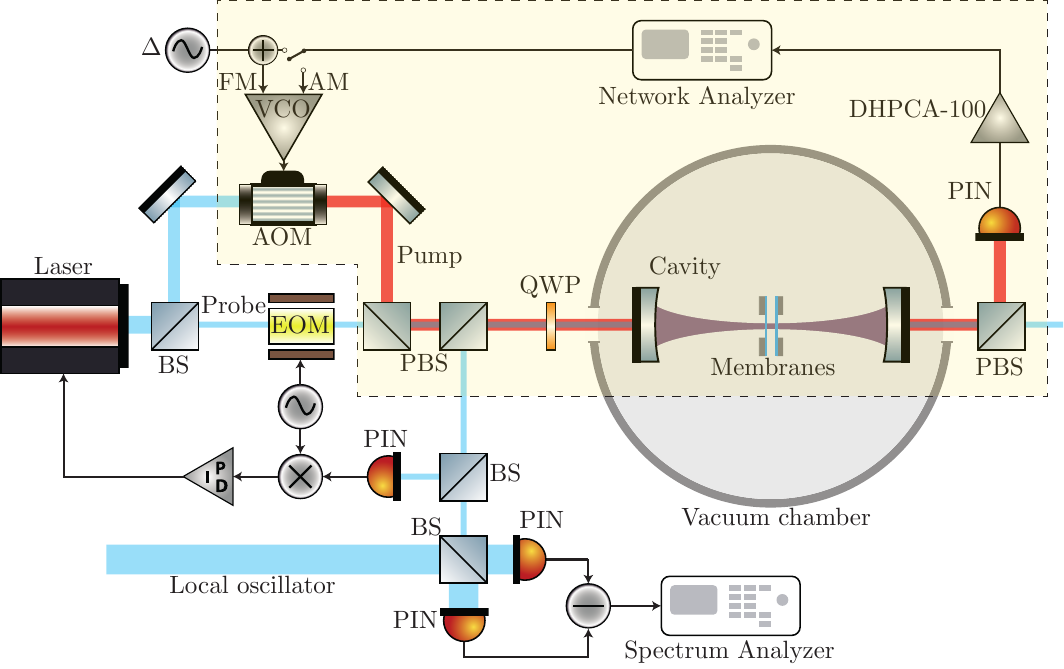}
\caption{Schematic of the experimental setup. The laser is split in a probe beam and a pump beam. The probe is modulated by an EOM. A fraction of the reflected component is measured to implement the PDH technique. The remaining field is mixed with a local oscillator to obtain the homodyne signal. The pump beam, whose path is highlighted by the yellow square, is modulated through an AOM. The AOM is driven by a VCO. The frequency modulation allows to detune in a controlled way the pump from the cavity resonance frequency. The input ports of the VCO allow to inject amplitude and phase noise into the system. The cavity transmitted light is detected directly by a PIN photodiode. The photocurrent is amplified with a FEMTO DHPCA-100 transimpedance amplifier.}
\label{fig:expsetup}
\end{center}
\end{figure}

The optomechanical setup is constituted by two Si$_3$N$_4$ square membranes within an optical cavity \cite{Piergentili2018, Piergentili2020, Piergentili2021, Piergentili:2022aa}.
A laser beam at wavelength $\lambda=\SI{1064}{\nano\meter}$ is split in a probe beam and a pump beam.
The first one has power $P_{pr}=\SI{3.8}{\micro\watt}$, and it is modulated by an electro-optical modulator (EOM).
It is locked to the cavity resonance frequency by means of Pound-Drever-Hall (PDH) technique.
The fraction of the beam reflected by the cavity is revealed by homodyne detection, yielding an effective measurement of the motion of the membranes.
The pump beam is more intense, and its power can be controlled. It is employed to realize the optomechanical interaction.
The pump beam is suitable also to inject amplitude and phase noise into the system and to scan the cavity linewidth.\\
\indent Amplitude and phase noise can be controlled by adding a modulation to the pump beam.
The light modulation is implemented through an acousto-optic modulator (AOM), driven by a voltage controlled oscillator (VCO).
We control the frequency and the amplitude of the modulation feeding the VCO with DC signals.
The frequency modulation (FM) is useful to detune the pump beam with respect to the cavity.
The amplitude and phase modulation is used to introduce a seed beam into the cavity.
We can evaluate the response function of the system to the amplitude and phase noise modulation, respectively, by measuring the light transmitted by the cavity at different frequencies of the seed.
A schematic view of the experimental setup is depicted in Fig.~\ref{fig:expsetup}.
The light transmitted by the cavity is collected on a PIN photodiode.
The photocurrent is then amplified by a FEMTO DHPCA-100 transimpedance amplifier with $\SI{1e6}{\volt/\ampere}$ gain and $\SI{3.5}{\mega\hertz}$ bandwidth.

\subsection{Experimental parameters}

\begin{figure}[htbp]
\begin{center}
\includegraphics[scale=0.83]{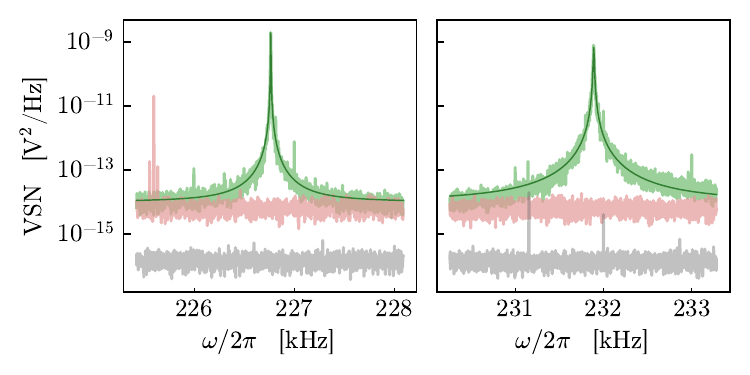}
\caption{Homodyne voltage spectral noise around mechanical fundamental modes frequencies. The light green curve shows the thermal noise of the mechanical oscillators. The mechanical parameters are estimated using the Lorentzian functions represented in green. The red curve indicates the shot noise level, the black one is the electronic noise of the device.}
\label{fig:mech_spect}
\end{center}
\end{figure}

The two membranes form themselves an inner cavity $L_c=\SI{53.571\pm0.009}{\micro\meter}$ long, the thickness of the Si$_3$N$_4$ layer is $L_m=\SI{106\pm1}{\nano\meter}$, and the transverse dimensions were estimated from the normal modes spectrum to be $L_x^{(1)}=\SI{1.519\pm0.006}{\milli\meter}$, $L_y^{(1)}=\SI{1.536\pm0.006}{\milli\meter}$, $L_x^{(2)}=\SI{1.522\pm0.006}{\milli\meter}$, $L_y^{(2)}=\SI{1.525\pm0.006}{\milli\meter}$.
A complete description of the characterization method is presented in \cite{Piergentili2018}.
We have studied the fundamental vibration mode of the two membranes by measuring the voltage spectral noise of the reflected cavity field, revealed by homodyne detection, as shown in Fig.~\ref{fig:mech_spect}.
The effective mass of the fundamental mode for both membranes is $m_{eff}=\SI{174}{\nano\gram}$.
The first oscillator is centered around $\omega_{m,1}=2\pi\times\SI{226.764581\pm0.000001}{\kilo\hertz}$, with a bandwidth (i.e., mechanical damping rate) $\gamma_{m,1}=2\pi\times\SI{1.44\pm0.01}{\hertz}$, yielding a mechanical quality factor $\mathcal{Q}_{m,1}=\SI{1.57e5}{}$.
For the second membrane we have found $\omega_{m,2}=2\pi\times\SI{231.88732\pm0.00002}{\kilo\hertz}$, $\gamma_{m,2}=2\pi\times\SI{8.8\pm0.1}{\hertz}$, and $\mathcal{Q}_{m,2}=\SI{2.63e4}{}$.
We evaluated the cavity decay rate by measuring its linewidth, and we extracted an amplitude decay rate $\kappa=2\pi\times\SI{119\pm1}{\kilo\hertz}$.
The detuning is fixed to $\Delta=2\pi\times\SI{240}{\kilo\hertz}$.
The experimental parameters are collected in Table~\ref{table:exp_pars}.

\begin{table}[h]
\begin{tabular}{|l|l|l|l|l|lll}
\cline{1-2} \cline{4-5} \cline{7-8}
               & Membrane 1                            &  &                & Membrane 2                            & \multicolumn{1}{l|}{} & \multicolumn{1}{l|}{}         & \multicolumn{1}{l|}{Cavity}                             \\ \cline{1-2} \cline{4-5} \cline{7-8}
$\omega_{m,1}$ & $2\pi\times\SI{226.764}{\kilo\hertz}$ &  & $\omega_{m,2}$ & $2\pi\times\SI{231.887}{\kilo\hertz}$ & \multicolumn{1}{l|}{} & \multicolumn{1}{l|}{$\kappa$} & \multicolumn{1}{l|}{$2\pi\times\SI{119}{\kilo\hertz}$} \\ \cline{1-2} \cline{4-5} \cline{7-8}
$\gamma_{m,1}$ & $2\pi\times\SI{1.44}{\hertz}$         &  & $\gamma_{m,2}$ & $2\pi\times\SI{8.8}{\hertz}$        & \multicolumn{1}{l|}{} & \multicolumn{1}{l|}{$\Delta$} & \multicolumn{1}{l|}{$2\pi\times\SI{240}{\kilo\hertz}$}  \\ \cline{1-2} \cline{4-5} \cline{7-8}
$m_{eff,1}$    & $\SI{174}{\nano\gram}$              &  & $m_{eff,2}$    & $\SI{174}{\nano\gram}$               &                       &                               &                                                         \\ \cline{1-2} \cline{4-5}
\end{tabular}
\caption{Optical and mechanical parameters.}
\label{table:exp_pars}
\end{table}

\subsection{Amplitude and phase noise calibration}

\begin{figure}[htbp]
\begin{center}
\includegraphics[scale=0.83]{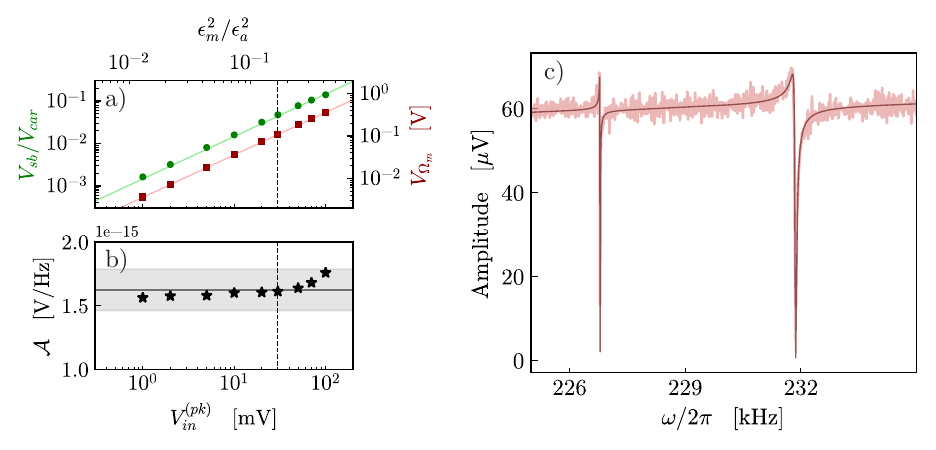}
\caption{Calibration of the noise and transmitted seed spectrum around the mechanical frequencies. a) The green dots represent the measured ratio between sideband and carrier amplitudes, while the red squares are the amplitude of the signal at frequency $\Omega_m$. The calibration is performed for different input voltage. b) Calculated detection factor for different seed amplitude. The horizontal line indicates the average of the points, with the shadow area which represents $\pm10\%$ of variation from the mean value. c) The seed is injected as amplitude modulation of the pump beam, using an input voltage $V_{in}^{(pk)}=\SI{30}{\milli\volt}$, highlighted by the vertical dashed line in panels a) and b). The component transmitted by the cavity is detected. The interference between the seed directly transmitted by the cavity, and the pump photons scattered by the oscillators, gives rise to the cancellations at the mechanical frequencies. The theoretical model, shown as a darker line on the experimental data, is calculated using the calibrated parameters. The single photon optomechanical couplings $g_{12}^{(0)}=2\pi\times\SI{0.13}{\hertz}$, $g_{22}^{(0)}=2\pi\times\SI{0.39}{\hertz}$, respectively, optimize the fitting.}
\label{fig:amp_noise_spect}
\end{center}
\end{figure}

We have experimentally verified the theoretical model by measuring the response of the optomechanical system to amplitude noise modulations on the pump beam.
The pump beam transmitted by the cavity, around the mechanical frequencies, is detected with the schematic presented in Fig.~\ref{fig:expsetup}.
To calibrate the quadrature spectrum we have to point out the relationship between the theoretical spectrum, i.e. the spectrum of Eq.~\eqref{X_out}, and the measured data.
Experimentally, a lock-in amplifier (LIA) is used to probe the response of the system at the modulation frequency $\Omega_m$.
We can calibrate the noise impinging on the cavity by measuring the ratio between the amplitude of the field sidebands, generated by the AOM modulation at frequency $\pm\Omega_m$, and the amplitude of the field at the carrier frequency.
An heterodyne scheme can be easily obtained by mixing the pump beam and the probe beam polarizations.
The amplitude of the pump field modulated by the AOM can be written as
\begin{equation}
\epsilon_{pu}(t)=\sqrt{\epsilon_a^2+\epsilon_m^2\cos(\Omega_m t)}e^{i(\omega_L-\Delta)t},
\end{equation}
where $\epsilon_a=\sqrt{P_{pu}/\hbar\omega_L}$, with $P_{pu}=\SI{67}{\micro\watt}$.
Instead, we do not add any artificial noise to the probe beam, and we can safely neglect the noise due to the laser (Coherent Mephisto 500). The field of the probe is
\begin{equation}
\epsilon_{pr}(t)=\epsilon_b e^{i\omega_L t}.
\end{equation}
The intensity of the light impinging on the photodiode is given by
\begin{align}
|\epsilon_{pu}+\epsilon_{pr}|^2&\simeq\mathrm{DC}+\epsilon_m^2\cos(\Omega_m t)+2\epsilon_a\epsilon_b\left[1+\frac{1}{2}\frac{\epsilon_m^2}{\epsilon_a^2}\cos(\Omega_m t)\right]\cos(\Delta t) \\
\nonumber			      &=\mathrm{DC}+\epsilon_m^2\cos(\Omega_m t)+2\epsilon_a\epsilon_b\cos(\Delta t)+2\epsilon_a\epsilon_b\frac{\epsilon_m^2}{4\epsilon_a^2}\left\{\cos\left[(\Omega_m+\Delta)t\right]+\cos\left[(\Omega_m-\Delta)t\right]\right\},
\end{align}
assuming a small modulation $\epsilon_m\ll\epsilon_a$. This form clearly reveals a DC signal, and oscillating components at $\Omega_m$, $\Delta$, $\Omega_m\pm\Delta$, respectively.
We can measure the amplitude of the signal at different frequencies by means of a LIA.
After the demodulation, the signal is passed through a fourth order band-pass filter with $\SI{19}{\hertz}$ bandwidth.
We perform the demodulation at frequencies $\Delta$, $\Omega_m+\Delta$, $\Omega_m$, to measure $V_{car}$, $V_{sb}$, and $V_{\Omega_m}$, respectively.
Eventually, we can calculate the amplitude of the modulation
\begin{equation}
\epsilon_m^2 = 4\epsilon_a^2\frac{V_{sb}}{V_{car}}=4\frac{P_{pu}}{\hbar\omega_L}\frac{V_{sb}}{V_{car}}.
\end{equation}
The calibration for different values of the modulation input voltage is presented in Fig.~\ref{fig:amp_noise_spect}a).
The interference between the modulated pump and the probe contains also a component at frequency $\Omega_m$, which is proportional to $\epsilon_m^2$.
The amplitude $V_{\Omega_m}$ is related to $\epsilon_m^2$ by
\begin{equation}
V_{\Omega_m}=\mathcal{A}\epsilon_m^2,
\end{equation}
where $\mathcal{A}$ takes into account the detection apparatus.
We underline that $V_{sb}$, and $V_{car}$ contain the detection term, too.
When we divide them to determine the ratio $\epsilon_m^2/\epsilon_a^2$ the factor $\mathcal{A}$ cancels out, but nonetheless it can be calculated from $V_{\Omega_m}$, $V_{sb}$, $V_{car}$ as follows:
\begin{equation}
\mathcal{A}=\frac{V_{\Omega_m}V_{car}}{4\epsilon_a^2V_{sb}}.
\end{equation}
The measured points for $V_{\Omega_m}$ at different modulation voltages are shown in Fig.~\ref{fig:amp_noise_spect}a), and the estimated $\mathcal{A}$ in the panel b) of the same figure.
Moreover we measured the seed transmitted by the cavity around the mechanical frequencies, injecting noise with voltage modulation $V_{in}^{(pk)}=\SI{30}{\milli\volt}$.
The measured data are shown in Fig.~\ref{fig:amp_noise_spect}c), and compared to the theoretical model given in Eq.~\eqref{X_out}.
The theoretical curves reproduce the measured data accurately, using a noise amplitude $\epsilon_m=\SI{8.2e6}{\Hz^{1/2}}$, and a detection factor $\mathcal{A}=\SI{1.6e-15}{\volt\per\hertz}$, both given by the calibration.
Experimentally, each point of the spectrum is obtained simulating the effect of a white noise within the bandwidth of the measurement $\mathrm{BW}=\SI{10}{\hertz}$, and the equivalent amplitude noise spectral density $\Gamma_{\epsilon}$, defined in Eq.~\eqref{eq:Gamma_noise} of the model, can be linked to $\epsilon_m$ through $2\Gamma_{\epsilon}=\epsilon_m^2/\mathrm{BW}\simeq\SI{6.7e12}{\hertz/\hertz}$.
The estimated value of the detection factor is valid for the measurement in Fig.~\ref{fig:amp_noise_spect}c).
In fact, the detection system used for the calibration, and the one employed to measure the light transmitted by the cavity, are analogous.
The best-fitting optomechanical couplings are comparable to what can be evaluated using the method of Ref. \cite{Piergentili2021} and are given by $g_{12}^{(0)}=2\pi\times\SI{0.13}{\hertz}$, $g_{22}^{(0)}=2\pi\times\SI{0.39}{\hertz}$.
The spectrum in Fig.~\ref{fig:amp_noise_spect}c) shows two Fano resonances with asymmetric shape, which is a typical manifestation of interference, e.g., in optomechanical systems it has been described in Refs.~\cite{Elste:2009aa, Qu:2013aa}.
In this case, the two dips in correspondence to the mechanical resonance frequencies of the membranes are due to the destructive interference between the amplitude noise of the driving pump directly transmitted by the cavity, and the optical output associated with the effective response of each mechanical resonator to the same amplitude noise.
In the experiment the two oscillators have different bare mechanical quality factors and optomechanical couplings, resulting in two distinct effective responses, and it appears in the spectrum as different shapes of the two dips.
This output field cancellation is similar to the optomechanically induced transparency (OMIT) \cite{Weis:2010aa, Karuza:2013aa}, in this case stimulated by the amplitude noise term.

\begin{figure}[htbp]
\begin{center}
\includegraphics[scale=0.83]{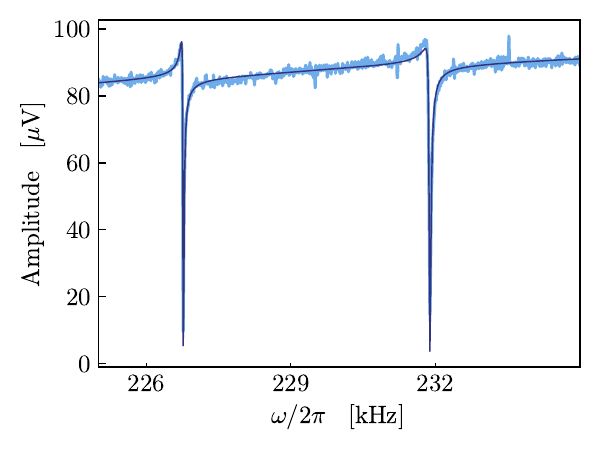}
\caption{The seed is injected as phase modulation of the pump beam and the output of the optical cavity is detected. The lighter curve represents the experimental data, with the best fitting function darker. Knowing the detection factor $\mathcal{A}=\SI{1.6e-15}{\volt\per\hertz}$ we can estimate the noise amplitude $\dot{\phi}=\SI{5.6e5}{\hertz}$. The single photon optomechanical couplings $g_{12}^{(0)}=2\pi\times\SI{0.42}{\hertz}$, $g_{22}^{(0)}=2\pi\times\SI{0.51}{\hertz}$, respectively, optimize the fitting.  The values of single photon optomechanical couplings are different from those of the amplitude noise case of the previous figure, because the membranes have been displaced along the cavity axis from the previous positions, yielding a different coupling situation, as illustrated in Ref.~\protect\cite{Piergentili2018}.}
\label{fig:phase_noise_spect}
\end{center}
\end{figure}

Alternatively, we can inject noise on the phase of the pump beam.
We have described above a method to evaluate the amount of amplitude noise on the beam, and we have introduced the factor $\mathcal{A}$ to take into account the detection.
Using the experimental apparatus represented in Fig.~\ref{fig:expsetup}, we have measured the cavity output also when phase noise is injected into the system.
In Fig.~\ref{fig:phase_noise_spect} we show the spectrum of the light intensity transmitted by the cavity, near to the mechanical resonances.
In this case the single photon optomechanical couplings are $g_{12}^{(0)}=2\pi\times\SI{0.42}{\hertz}$, and $g_{22}^{(0)}=2\pi\times\SI{0.51}{\hertz}$.
The value of the detection factor is already determined, and it is given by $\mathcal{A}=\SI{1.6e-15}{\volt\per\hertz}$.
We can estimate the intensity of the phase noise seed by fitting the amplitude of the experimental data, and we get $\dot{\phi}=\SI{5.6e5}{\hertz}$.
This result yields the equivalent phase noise spectral density $2\Gamma_{L}=\dot{\phi}^2/BW\simeq\SI{3.1e10}{\hertz^2/\hertz}$, which can be intended as the linewidth of the noisy simulated laser.
The two dips in Fig.~\ref{fig:phase_noise_spect} are addressed again to an OMIT-like behavior, here stimulated by the phase noise seed.

\subsection{Noise cancellation}

\begin{figure}[htbp]
\begin{center}
\includegraphics[scale=0.83]{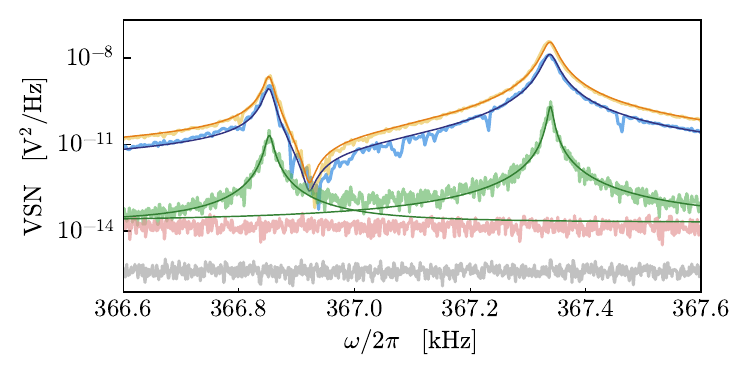}
\caption{Homodyne voltage spectral noise around the first excited mechanical mode frequencies of one membrane. The lighter curves represent the measured data, the solid darker lines the theoretical best-fitting functions. The green curve represents the thermal noise of the membrane, i.e. the homodyne signal spectrum when the pump beam is switched off. We estimate the mechanical parameters by fitting it with Lorentzian peaks. The orange and blue lines are measured injecting a seed of intensity $\epsilon_m^2=\SI{6.7e13}{\hertz}$ and $\epsilon_m^2=\SI{1.1e14}{\hertz}$, respectively, through the pump beam amplitude. Sweeping the frequency of the seed we get the spectra. A noise cancellation window can be noted within the two thermal peaks frequencies. The shot noise is presented in red, and the electronic noise is gray.}
\label{fig:amp_noise_cancellation}
\end{center}
\end{figure}

Eventually, we can detect the mechanical displacement noise by means of homodyne detection of the reflected probe beam, while injecting amplitude noise in the system through the pump beam.
The detection reveals the phase quadrature of the probe field, in presence of amplitude noise on the pump.
The same measurement can be performed also adding phase noise, with almost identical results.
The noise calibration described in the previous section remains valid.
The first excited modes of one membrane, labeled $(12)$, $(21)$, have resonance frequencies $\omega_m^{(12)}=\SI{366.8525\pm0.0002}{\kilo\hertz}$, $\omega_m^{(21)}=\SI{367.3389\pm0.0002}{\kilo\hertz}$, and mechanical damping rates $\gamma_m^{(12)}=\SI{11.9\pm0.5}{\hertz}$, $\gamma_m^{(21)}=\SI{8.6\pm0.4}{\hertz}$, respectively.
The characterization of the mechanical parameters has been performed by fitting the mechanical displacement spectrum with Lorentzian peaks, as shown in Fig.~\ref{fig:amp_noise_cancellation}.
Closer resonance frequencies than the fundamental modes allow to observe easier a cancellation window within them.
A cancellation of the amplitude noise is observed, and the theoretical result in Eq.~\eqref{Y_out} describes the measured data with excellent agreement.
The orange and blue curves in Fig.~\ref{fig:amp_noise_cancellation} refer to the case when amplitude noise is much larger than the other noise sources.
As a consequence, the two mechanical modes are excited via radiation pressure by the same fluctuating force.
In the spectral region between the resonances the two mechanical responses are out of phase and destructively interfere, yielding a visible noise cancellation.
Similar cancellation figures have been observed also in different experimental setup, and they can be engineered to improve the sensitivity of the measurement within the cancellation's bandwidth \cite{Caniard:2007aa, Moaddel-Haghighi:2018aa}.

\section{Conclusion}

In this work we considered a multimode cavity optomechanical system in which two SiN membranes are placed within a Fabry-Per\'ot cavity and driven by two laser fields.
We provided a description of the effects of the laser's amplitude and phase noises, which are introduced in the system's equations of dynamics as an additive term, and a multiplicative term, respectively.
Moreover, we have used an artificial source of white noise to prove experimentally the validity of the model.
We evaluated the equivalent noise spectral density when the artificial noises overwhelm the other sources of noise.
Finally, the effective displacement spectral noise of the membranes in presence of amplitude noise has been measured: a cancellation window within the mechanical resonance frequencies arises due to the opposite sign of the phases in the mechanical response.
The theoretical model with the calibrated value of the noise provided an accurate description also in this case.

\section*{Acknowledgments}

We acknowledge financial support from NQSTI within PNRR MUR project PE0000023-NQSTI.

\bibliographystyle{unsrt} 
\bibliography{amp_noise_bib}

\end{document}